\input{aipcheck.tex}

\newcommand\sps{\space\space\space\space}
 \def\selectedoptions{final}

\documentclass[
   \selectedoptions
  ]
  {aipproc}

\def\selectedlayoutstyle{8x11double}
\layoutstyle\selectedlayoutstyle
\SetInternalRegister\hbadness{8000} 

\newcommand\doingARLO[2][]{%
  \ifx\mmref\undefined #1\else #2\fi
}

\begin{document}

\title{Neutrino Interaction Calculations from MeV to GeV Region }


\author{J.E.~Amaro}{
  address={Departamento de
 F{\'\i}sica At\'omica, Molecular y Nuclear,\\ Universidad de Granada,
 E-18071 Granada, Spain},
  email={jmnieves@ugr.es},
}
\iftrue

\author{J.~Nieves}{
  address={Departamento de
 F{\'\i}sica At\'omica, Molecular y Nuclear,\\ Universidad de Granada,
 E-18071 Granada, Spain},
  email={jmnieves@ugr.es},
}

\author{M.~Valverde}{
  address={Departamento de
 F{\'\i}sica At\'omica, Molecular y Nuclear,\\ Universidad de Granada,
 E-18071 Granada, Spain},
  email={mvalverd@ugr.es}}

\author{M.J.~Vicente-Vacas}{
  address={Departamento de
F\'\i sica Te\'orica and IFIC, Centro Mixto Universidad de
Valencia-CSIC\\ Institutos de Investigaci\'on de Paterna,
Aptdo. 22085, E-46071 Valencia, Spain },}

\fi

\copyrightyear  {2001}

\begin{abstract}
  The Quasi-Elastic (QE) contribution of the nuclear inclusive
  electron model developed in reference~\cite{GNO97} is extended to
  the study of neutrino/antineutrino Charged Current (CC) and Neutral
  Current (NC) induced nuclear reactions at intermediate energies.
  Long range nuclear (RPA) correlations, Final State Interaction (FSI)
  and Coulomb corrections are included within the model. RPA
  correlations are shown to play a crucial role in the whole range
  (100--500 MeV) of studied neutrino   energies.  Results for inclusive
  muon capture for different nuclei through the Periodic Table are
  also discussed. In addition, and by means of a Monte Carlo cascade
  method to account for the rescattering of the outgoing nucleon, we
  also study the CC and NC inclusive one nucleon knockout reactions
  off nuclei.
\end{abstract}

\date{\today}

\maketitle

\section{Introduction}
Neutrino physics is at the forefront of current theoretical and
experimental research in astro, nuclear, and particle physics. Indeed,
neutrino interactions offer unique opportunities for exploring
fundamental questions in these domains of the physics. One of these
questions is the neutrino-oscillation phenomenon, for which there have
been conclusive positive signals in the last years~\cite{osc}. The
presence of neutrinos, being chargeless particles, can only be
inferred by detecting the secondary particles they create when
colliding and interacting with matter. Nuclei are often used as
neutrino detectors, thus a trustable interpretation of neutrino data
heavily relies on detailed and quantitative knowledge of the features
of the neutrino-nucleus interaction. There is a general consensus
among the theorists that a simple Fermi Gas (FG) model, widely used in
the analysis of neutrino oscillation experiments, fails to provide a
satisfactory description of the measured cross sections, and inclusion
of further nuclear effects is needed~\cite{sakuda}.

Any model aiming at describing the interaction of neutrinos with
nuclei should be firstly tested against the existing data on the
interaction of real and virtual photons with nuclei. For nuclear
excitation energies ranging from about 100 MeV to 500 or 600 MeV,
three different contributions should be taken into account: i) QE
processes, ii) pion production and two body processes from the QE
region to that beyond the $\Delta(1232)$ resonance peak, and iii)
double pion production and higher nucleon resonance degrees of freedom
induced processes.  The model developed in Refs.~\cite{GNO97}
(inclusive electro--nuclear reactions) and~\cite{CO92} (inclusive
photo--nuclear reactions) has been successfully compared with data at
intermediate energies and it systematically includes the three type of
contributions mentioned above.  Nuclear effects are computed starting
from a local FG picture of the nucleus, an accurate
approximation to deal with inclusive processes which explore the whole
nuclear volume~\cite{CO92}, and their main features, expansion
parameter and all sort of constants are completely fixed from previous
hadron-nucleus studies (pionic atoms, elastic and inelastic
pion-nucleus reactions, $\Lambda-$ hypernuclei,
etc...)~\cite{OTW82,pion}.  Thus, and besides the photon coupling
constants determined in the vacuum, the model of Refs.~\cite{GNO97}
and \cite{CO92} has no free parameters, and thus the results presented
in these two references are predictions deduced from the nuclear
microscopic framework developed in Refs.~\cite{OTW82} and \cite{pion}.
In this talk, we extend the nuclear inclusive QE electron scattering
model of Ref.~\cite{GNO97}, including the axial CC~\cite{ccjuan} and
NC~\cite{ncjuan} degrees of freedom, to describe neutrino and
antineutrino induced nuclear reactions in the QE region. 

\begin{figure}
\includegraphics[height=3.8cm]{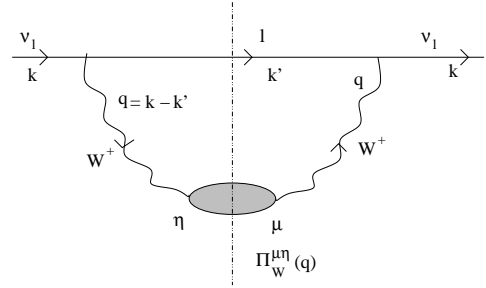}
  \caption{Diagrammatic representation of the neutrino
  selfenergy in nuclear matter.}
\label{fig:fig1}
\end{figure}
\begin{figure}
\includegraphics[height=6.3cm,width=6cm]{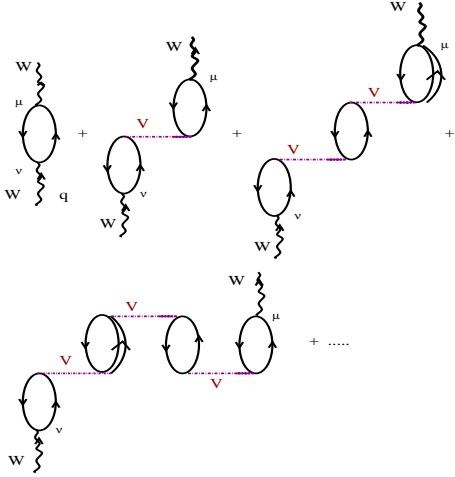}
\vspace{-0.8cm}
 \caption{\footnotesize Set of irreducible diagrams responsible for the
  polarization (RPA) effects in the 1p1h contribution to the
  $W-$selfenergy. }\label{fig:fig3}
\end{figure}
\begin{figure}[h]
 \vspace{-2.8cm}
 \includegraphics[width=6.7cm,height=8cm]{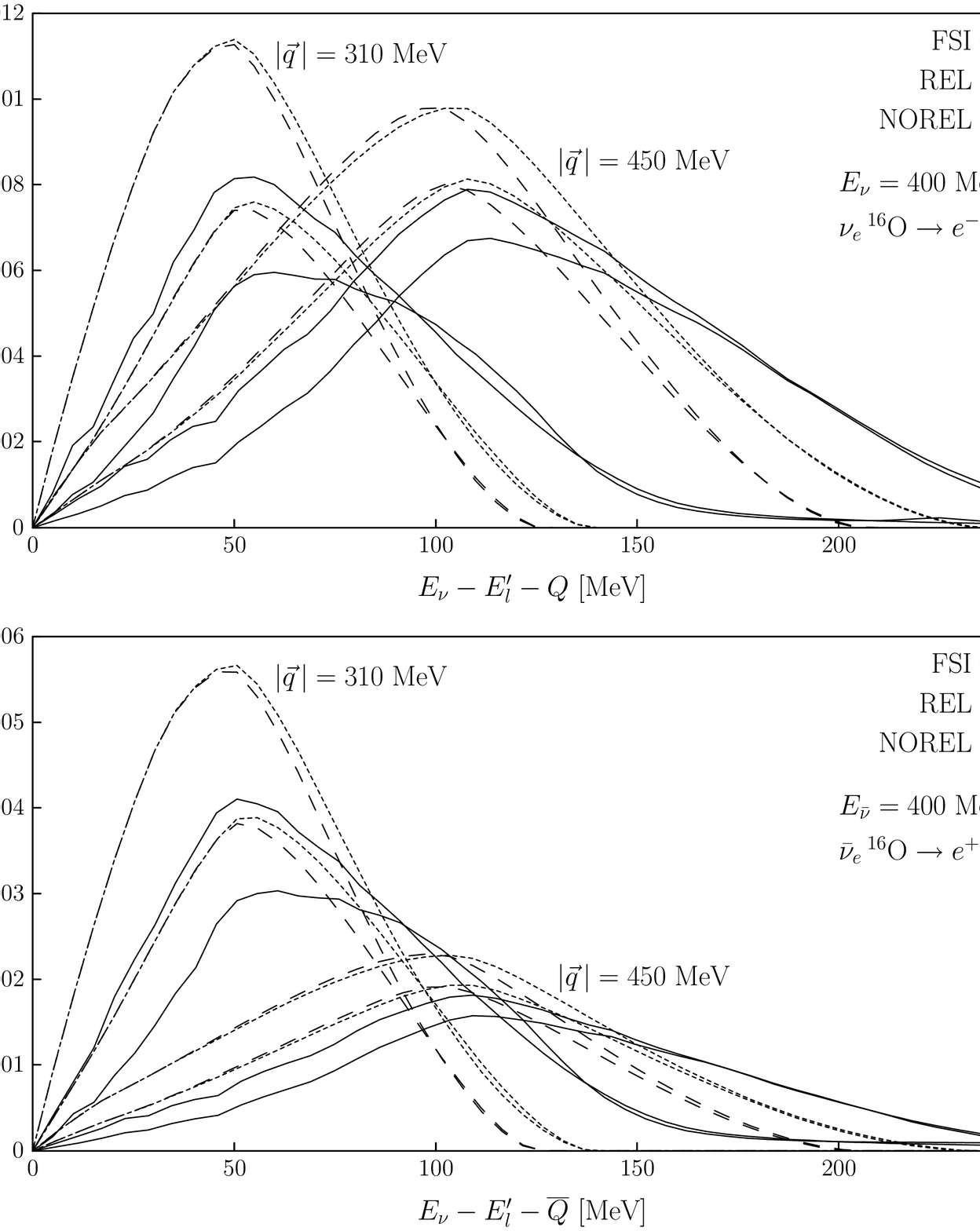}
\vspace{-1.7cm}
\caption{\footnotesize $\nu_e-$ (top) and ${\bar \nu}_e-$ (bottom)
  inclusive QE differential cross sections in oxygen as a function of
  the transferred energy, at two values of the transferred momentum. We show
  results for relativistic (REL) and non-relativistic nucleon
  kinematics. In this latter case, we present results with (FSI) and
  without (NOREL) FSI effects. For the three cases, we also show the
  effect of taking into account RPA correlations and Coulomb
  corrections (lower lines at the peak).}
  \label{fig:fsi}
\end{figure}
\begin{figure}
\vspace{-0.9cm}
\includegraphics[width=6.5cm,height=4cm]{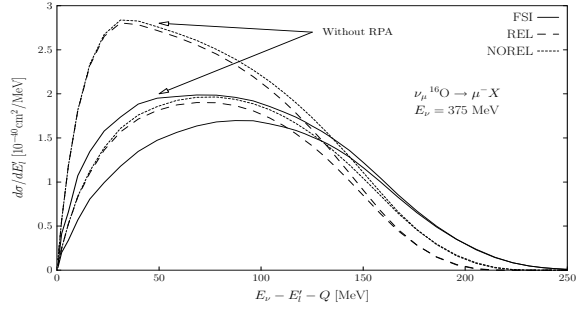}
\caption{ \footnotesize Muon neutrino inclusive QE differential cross
  sections in oxygen as a function of the transferred energy.  The
  notation for the theoretical predictions is the same as in
  Fig.~\protect\ref{fig:fsi}.  }
  \label{fig:fsi3}
\end{figure}
We also present results for the QE $(\nu_l,\nu_l N)$, $(\nu_l,l^- N)$,
$({\bar \nu}_l,{\bar \nu}_l N)$ and $({\bar \nu}_l,l^+ N)$ reactions
in nuclei. We use a Monte Carlo (MC) simulation method to account for
the rescattering of the outgoing nucleon.  The first step is the gauge
boson ($W^{\pm}$ and $Z^0$ ) absorption in the nucleus. We take this
reaction probability from the microscopical many body framework
developed in Refs.~\cite{ccjuan,ncjuan} for CC and NC induced
reactions.  Some calculations found in the literature use the plane
wave and distorted wave impulse approximations (PWIA and DWIA,
respectively), including or not relativistic effects.  The PWIA
constitutes a poor approximation, since it neglects all types of
interactions between the ejected nucleon and the residual nuclear
system. The DWIA describes the ejected nucleon as a solution of the
Dirac or Schr\"odinger equation with an optical potential obtained by
fitting elastic proton--nucleus scattering data. The imaginary part
accounts for the absorption into unobserved channels. This scheme is
incorrect to study nucleon emission processes where the state of the
final nucleus is totally unobserved, and thus all final nuclear
configurations, either in the discrete or on the continuum,
contribute.  The distortion of the nucleon wave function by a complex
optical potential removes all events where the nucleons collide
inelastically with other nucleons. Thus, in DWIA calculations, the
nucleons that interact inelastically are lost when in the physical
process they simply come off the nucleus with a different energy,
angle, and maybe charge, and they should definitely be taken into
account. A clear example which illustrates the deficiencies of the
DWIA models is the neutron emission process: $(\nu_l, l^- n)$. Within
the impulse approximation neutrinos only interact via CC interactions
with neutrons and would emit protons, and therefore the DWIA will
predict zero cross sections for CC one neutron knock-out reactions.
However, the primary protons interact strongly with the medium and
collide with other nucleons which are also ejected. As a consequence
there is a reduction of the flux of high energy protons but a large
number of secondary nucleons, many of them neutrons, of lower energies
appear.

Finally, we would like to mention that the we have already started
working on one and two pion production processes in nuclei, aiming to
extend the present study beyond the QE peak to the $\Delta(1232)$
resonance region. As a first step, we have studied the production off
the nucleon~\cite{prd,twopion} and some results have been also
reported at this conference~\cite{nufact-pion}.

\section{Inclusive QE cross sections}

We will expose here the general formalism focusing on the neutrino CC
reaction. The generalization to antineutrino CC, neutrino and
antineutrino NC reactions or inclusive muon capture is straightforward.
In the laboratory frame, the differential cross section for the
process $\nu_l (k) +\, A_Z \to l^- (k^\prime) + X $ reads:
\begin{equation}
\frac{d^2\sigma}{d\Omega(\hat{k^\prime})dE^\prime_l} =
\frac{|\vec{k}^\prime|}{|\vec{k}~|}\frac{G^2}{4\pi^2} 
L_{\mu\sigma}W^{\mu\sigma} 
\end{equation}
with $L$ and $W$ the leptonic and hadronic tensors, respectively. On
the other hand, the inclusive CC nuclear cross section is related to
the imaginary part of the neutrino self-energy (see
Fig.~\ref{fig:fig1}) in the medium by:
\begin{equation}
\sigma = - \frac{1}{|\vec{k}|} \int {\rm Im} \Sigma_\nu (k;\rho(r)) d^3r
\label{eq:sSrel}
\end{equation}
We obtain the imaginary part of the neutrino self-energy in the
medium, ${\rm Im} \Sigma_\nu$, by means of the Cutkosky's rules: in
this case we cut with a vertical straight line (see
Fig.~\ref{fig:fig1}) the intermediate lepton state and those implied
by the $W$-boson medium self-energy.  Those states are placed on shell
by taking the imaginary part of the propagator, self-energy, etc. We
obtain for $k^0 > 0$
\begin{equation}
  {\rm Im} \Sigma_\nu(k) = \frac{8G\,\Theta(q^0)}{\sqrt 2 M^2_W}\int \frac{d^3
  k^\prime}{(2\pi)^3 }\frac{ {\rm Im}\left\{ \Pi^{\mu\eta}_W 
L_{\eta\mu} \right\}}{2E^{\prime}_l}  
\label{eq:ims}
\end{equation}
and thus, the hadronic tensor is basically an integral over the
nuclear volume of the $W-$selfenergy ($\Pi_W^{\mu\nu}(q;\rho)$) inside
the nuclear medium. We can then take into account the different
in-medium effects and reaction mechanism modes ($W$ absorption by one
nucleon or by a pair of nucleons, pion production, excitation of
resonances,...) by including the correspondent diagrams in
the $W-$selfenergy (shaded loop of Fig.~\ref{fig:fig1}). Further details
can be found in Refs.~\cite{GNO97,ccjuan}.

The virtual $W$ gauge boson can be absorbed by one nucleon, 1p1h
nuclear excitation, leading to the QE contribution to the nuclear response
function.  We consider a structure
of the $V-A$ type for the $W^+pn$ vertex, and use PCAC and invariance
under G-parity to relate the pseudoscalar form factor to the axial one
and to discard a term of the form $(p^\mu+p^{\prime \mu})\gamma_5$ in
the axial sector, respectively.   Besides, and thanks to
isospin symmetry, the vector form factors are related to the
electromagnetic ones.  We find
\begin{eqnarray}
W^{\mu\nu}(q) 
&=&\frac{\cos^2\theta_C}{2M^2} \int_0^\infty dr r^2 \Big \{
\Theta(q^0) \int \frac{d^3p}{4\pi^2} \frac{M}{E(\vec{p})}
 \nonumber \\
&& \hspace{-2cm}\frac{M}{E(\vec{p}+\vec{q})}  
 \Theta(k_F^n(r)-|\vec{p}~|) \Theta(|\vec{p}+\vec{q}~|-k_F^p(r))
 \nonumber\\
 & &\hspace{-2cm} \delta\left(q^0 +
E(\vec{p}) -E(\vec{p}+\vec{q}~)\right)   A^{\nu\mu}(p,q)|_{p^0=E(\vec{p})~} 
\Big \}
\label{eq:thadron}
\end{eqnarray}
with the local Fermi momentum $k_F(r)= (3\pi^2\rho(r)/2)^{1/3}$, $M$
the nucleon mass, and $E(\vec{p}\,)= \sqrt{M^2 + \vec{p}^{\,2}}$. We
work on an non-symmetric nuclear matter with different Fermi sea
levels for protons, $k_F^p$, than for neutrons, $k_F^n$ (equation
above, but replacing $\rho /2$ by $\rho_p $ or $\rho_n$, with
$\rho=\rho_p + \rho_n$). Finally, $A^{\mu\nu}$ is the CC nucleon
tensor~\cite{ccjuan} and is determined by the $W^+pn$ form-factors.
The $d^3p$ integrations above can be done analytically and all of them
are determined by the imaginary part of isospin asymmetric Lindhard
function, $\overline {U}(q,k_F^n,k_F^p)$.  Explicit expressions can be
found in ~\cite{ccjuan}, where also antineutrino induced CC cross
sections and the inclusive muon capture process in nuclei are
discussed. Expressions for the NC hadron tensor can be found in
Ref.~\cite{ncjuan}.

\subsection{Nuclear Corrections}
We take into account polarization effects by substituting the
particle-hole (1p1h) response by an RPA response consisting of a
series of ph and $\Delta$-h excitations, as shown in
Fig.~\ref{fig:fig3}.  We use an effective Landau-Migdal ph-ph
interaction~\cite{Sp77}: $V = c_{0}\left\{
  f_{0}+f_{0}^{\prime}\vec{\tau}_{1}\vec{\tau}_{2}+
  g_{0}\vec{\sigma}_{1}\vec{\sigma}_{2}+g_{0}^{\prime}
  \vec{\sigma}_{1}\vec{\sigma}_{2} \vec{\tau}_{1}\vec{\tau}_{2}
\right\}$.  In the vector-isovector channel ($\vec{\sigma} \vec{\sigma}
\vec{\tau} \vec{\tau}$ operator) we use an
interaction~\cite{GNO97,CO92,pion} with explicit $\pi-$meson
(longitudinal) and $\rho-$meson (transverse) exchanges, and that also
includes $\Delta(1232)$ degrees of freedom.  This effective
interaction is non-relativistic, and then for consistency we will
neglect terms of order ${\cal O}(p^2/M^2)$ when summing up the RPA
series. 

We also ensure the correct energy balance of the different
studied processes by modifying the energy conserving $\delta$ function
in Eq.~(\ref{eq:thadron}) to account for the experimental $Q-$value of
the reaction. Besides, we consider the effect of the Coulomb field of
the nucleus acting on the ejected charged lepton, as well. This is
done by including the lepton self- energy $\Sigma_C=2k^{\prime
  \,0}V_C(r)$ in the intermediate lepton propagator of Fig.
\ref{fig:fig1}. 

Finally, we take into account the modification of
nucleon dispersion relation in the medium (FSI) by using nucleon
propagators properly dressed with a realistic self-energy. We use a
dynamical model in which the nucleon self-energy depends explicitly on
the energy and the momentum~\cite{FO92}.  Thus, we compute the
imaginary part of the Lindhard function (ph propagator) using
realistic spectral functions $S_{p,h}(\omega,\vec p;\rho)$.
\begin{figure}
\includegraphics[width=7.7cm,height=4.5cm]{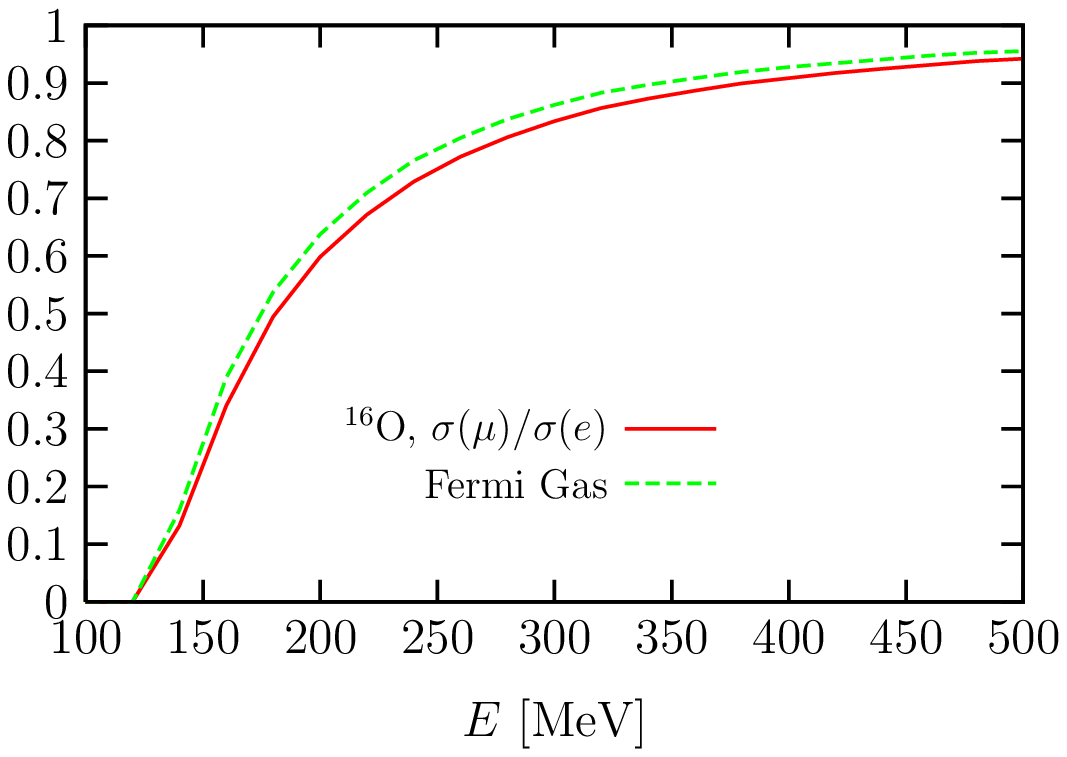}
\end{figure}
\begin{figure}
\includegraphics[width=7.7cm,height=4.5cm]{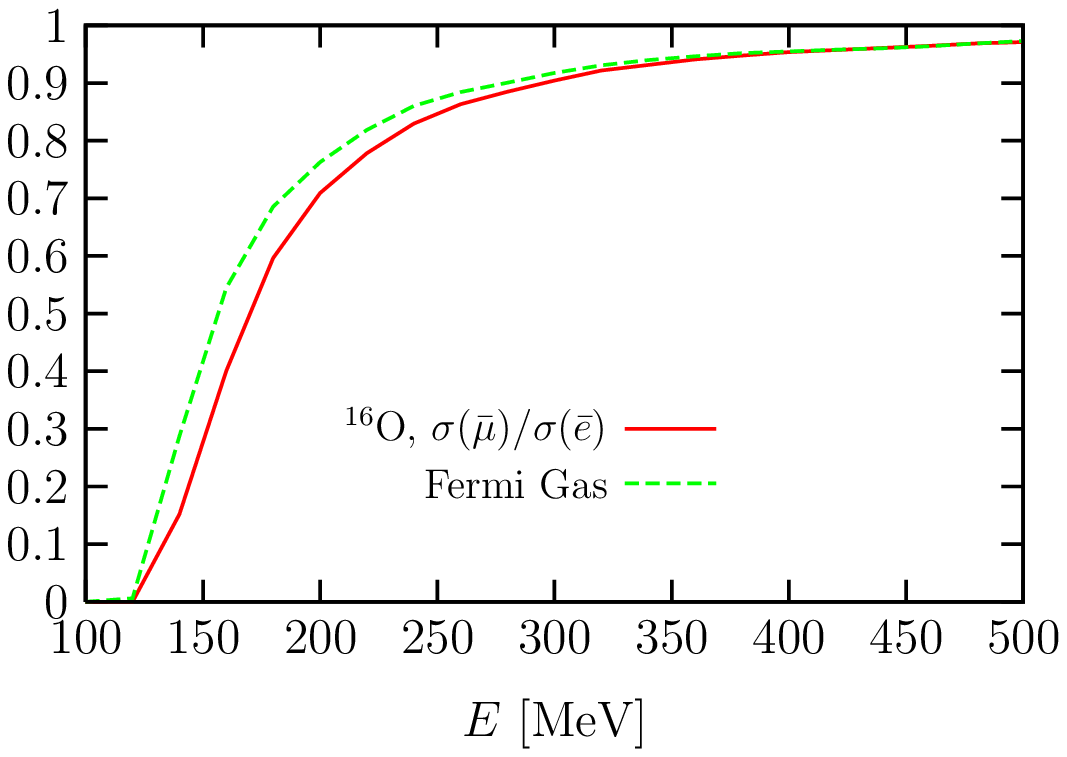}
\caption{\footnotesize Ratio of inclusive QE cross sections
 induce  by neutrinos (top) and antineutrinos (bottom) in oxygen, as a
  function of the incoming neutrino energy. Solid and dashed lines stand
for the predictions of the model of Ref.~\cite{ccjuan} and from the FG
model, respectively. }
  \label{fig:ratios}
\end{figure}
\section{CC and NC inclusive one nucleon knockout reactions}

We use a cascade method to account for the rescattering of the
outgoing nucleon.  The first step is the gauge boson ($W^{\pm}$ and
$Z^0$ ) absorption in the nucleus, we take this reaction probability
from the microscopical many body framework outlined in the previous
section for CC and NC induced reactions. It is given by the inclusive
QE cross sections $d^2\sigma /d\Omega^{\prime} dE^{\prime}$
($\Omega^{\prime}$, $E^{\prime}$ are the solid angle and energy of the
outgoing lepton) for a fixed incoming neutrino or antineutrino
laboratory  energy. We also compute differential cross sections
with respect to $d^3r$. Thus, we also know the point of the nucleus
where the gauge boson was absorbed, and we can start from there our MC
propagation of the ejected nucleon. After the absorption of the gauge
boson, we follow the path of the ejected nucleon through its way out
of the nucleus using a MC simulation  to account for the
secondary collisions. Details on the MC simulation can be found
in~\cite{ncjuan}. This MC simulator  has been
tested in different physical situations. It was first designed for
single and multiple nucleon and pion emission reactions induced by
pions~\cite{pion1,VicenteVacas:1993bk} and has been successfully
employed to describe inclusive $(\gamma, \pi)$, $(\gamma, N)$,
$(\gamma, NN)$,..., $(\gamma, N\pi)$,...~\cite{CO92bis,COV94}, $(e,
e^\prime \pi)$, $(e, e^\prime N)$, $(e, e^\prime NN)$,..., $(e,
e^\prime N\pi)$,...~\cite{GNO97b} reactions in nuclei or the neutron
and proton spectra from the decay of $\Lambda$
hypernuclei~\cite{Ramos:1996ik}.

\section{Results}

\subsection{Inclusive cross sections}

At low energies, the model provides a reasonable description of the
LSND measurement of the reaction $^{12}$C $(\nu_\mu,\mu^-)X$ near
threshold, and of the accurate nuclear inclusive muon capture rates
through the whole Periodic Table. Pauli blocking and the use of the
correct energy balance provide the bulk of the corrections, but an
accurate description of data is only achieved once the RPA
correlations, including $\Delta h$ excitations, are taken into
account. Our approach provides one of the best existing combined
descriptions of the inclusive muon capture in $^{12}$C and the LSND
measurement of the reaction $^{12}$C $(\nu_\mu,\mu^-)X$ near
threshold~\cite{ccjuan}, and certainly comparable to that achieved by
other models, which implement a more sophisticated treatment of the
dynamics of finite nuclei (see for instance the discussion
in~\cite{Ko03}).

At intermediate energies the predictions of our model become reliable
not only for integrated, but also for differential cross sections. We
present results for incoming neutrino energies within the interval
150-500 MeV for electron and muon species (some results at higher
energies can be found in \cite{Athar:2007wd}).  In Figs.~\ref{fig:fsi}
and~\ref{fig:fsi3}, RPA and FSI effects on differential cross section
are shown.  RPA effects are extremely important in this range of
energies and induce important corrections to the simple FG
description. On the other hand, FSI provides a
broadening and a significant reduction of the strength of the QE peak.
Nevertheless the integrated cross section is only slightly modified.
Though FSI change importantly the shape of the differential cross
sections, it plays a minor role for totally integrated cross sections.
When medium polarization effects are not considered, FSI provides
significant reductions (15-30\%) of the cross sections. However, when
RPA corrections are included the reductions becomes more moderate,
always smaller than 7\%, and even there exist some cases where FSI
enhances the cross sections. This can be easily understood by looking
at Fig.~\ref{fig:fsi3}. There, we see that FSI increases the cross
section at high energy transfers. But for nuclear excitation energies
higher than those around the QE peak, the RPA corrections are
certainly less important than in the peak region. Thus, the RPA
suppression of the FSI distribution is significantly smaller than the
RPA reduction of the distribution determined by the ordinary Lindhard
function.

We have estimated the theoretical uncertainties of our model by MC
propagating the uncertainties of its different inputs into
differential and total cross sections~\cite{ccjuan-errors}.  We
conclude that our approach provides QE $\nu(\bar \nu)$--nucleus cross
sections with relative errors of about
10-15\%, while uncertainties affecting the ratios
$\sigma(\mu)/\sigma(e)$ and $\sigma(\bar\mu)/\sigma(\bar e)$ would be
certainly smaller, not larger than about 5\%, and mostly coming
from deficiencies of the local FG picture of the
nucleus~\cite{ccjuan-errors}. Though nuclear corrections cancel
importantly in these ratios, there still exist some effects (
Fig.~\ref{fig:ratios}). 

\subsection{Nucleon Emission Reactions}

CC and NC nucleon emission processes play an important role in the
analysis of oscillation experiments. In particular, they constitute
the unique signal for NC neutrino driven reactions. Different
distributions for both NC and CC processes can be found in
~\cite{ncjuan}, as example, we show here results for NC nucleon
emission from argon (Fig.~\ref{fig:res5}). The rescattering of the
outgoing nucleon produces a depletion of the high energy side of
the spectrum, but the scattered nucleons clearly enhance the low
energy region. Our results compare well with those of
Refs.~\cite{Buss:2007ar, Leitner:2006sp} 
obtained by means of a transport model.
\begin{figure}[th]
\includegraphics[width=7.7cm,height=4.5cm]{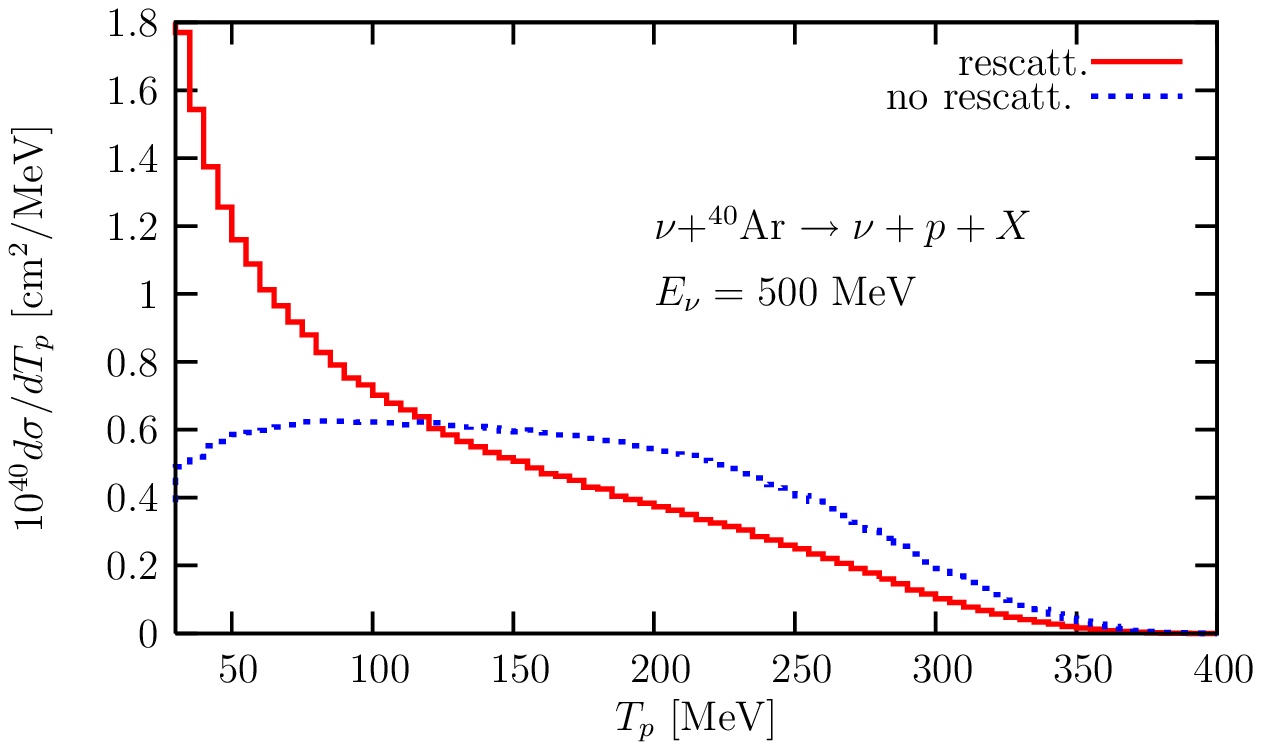}
\end{figure}
\begin{figure}[h]
\includegraphics[width=7.7cm,height=4.5cm]{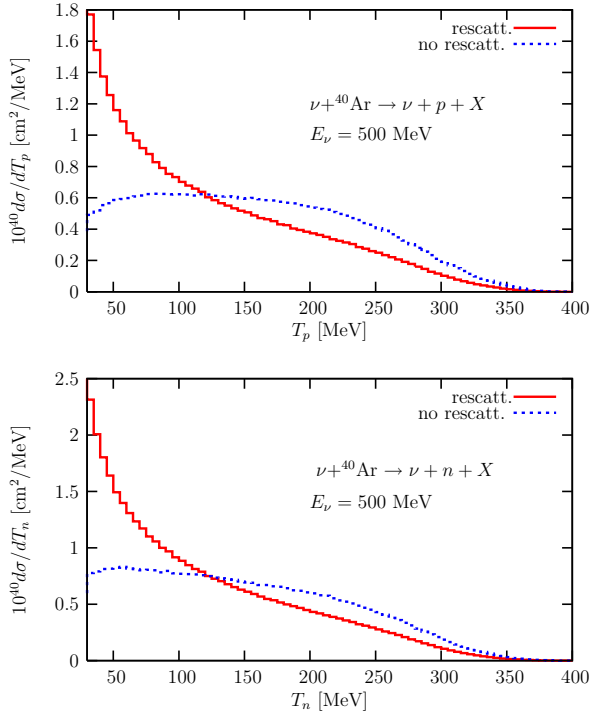}
\caption{ Neutral current $^{40}Ar(\nu,\nu+N)$ at 500 MeV cross
  sections as a function of the kinetic energy of the final nucleon.
  The dashed histograms show results without rescattering (PWIA) and
  the solid ones have been obtained from the MC cascade simulation.
}\label{fig:res5}
\end{figure}
\begin{theacknowledgments}
This work was partially supported by MEC contracts 
FIS2005-00810, FIS2006-03438, by the Generalitat Valenciana and by
Junta de Andaluc\'\i a under contracts
ACOMP07/302 and FQM0225, and by the EU under contract
RII3-CT-2004-506078.
\end{theacknowledgments}

\bibliographystyle{aipproc}   

\end{document}